# Using globular clusters to test gravity in the weak acceleration regime

Riccardo Scarpa[1], Gianni Marconi[2], Roberto Gilmozzi[2], and Giovanni Carraro[3]

[1]Instituto de Astrofísica de Canarias, Spain
[2]European Southern Observatory
[3]Universita' di Padova, Italy

We report on the results from an ongoing program aimed at testing Newton's law of gravity in the low acceleration regime using globular clusters. It is shown that all clusters studied so far do behave like galaxies, that is, their velocity dispersion profile flattens out at large radii where the acceleration of gravity goes below $10^{-8}$ cm s$^{-2}$, instead of following the expected Keplerian fall off. In galaxies this behavior is ascribed to the existence of a dark matter halo. Globular clusters, however, do not contain dark matter, hence this result might indicate that our present understanding of gravity in the weak regime of accelerations is incomplete and somehow incorrect.

Stars within galaxies, and galaxies within clusters of galaxies are very far apart from each other. As a consequence, the typical accelerations governing the dynamics of galaxies are orders of magnitude smaller than the ones probed in our laboratories or in the solar system. Thus, any time Newton's law is applied to galaxies (e.g., to infer the existence of dark matter), its validity is severely extrapolated. Although there are in principle no reasons to distrust Newton's law in the weak acceleration regime, unanimous agreement has been reached (e.g., Binney 2004) on the fact that galaxies start deviating from Newtonian dynamics, and dark matter is needed to reconcile observations with predictions, always for the same value of the internal acceleration of gravity $a_0 \sim 10^{-8}$ cm s$^{-2}$. This systematics, more than anything else, suggests that we may be facing a breakdown of Newton's law rather than the effects of dark matter.

Being free falling toward the Milky Way, the internal dynamics of globular clusters is only affected by tidal stress, which is in most cases well below $a_0$. Therefore the internal dynamics of globular clusters can be used to probe the same range of accelerations probed by galaxies without the complication of dark matter.

According to this idea we studied the dynamics of the external regions of ω Centauri (Scarpa, Marconi, and Gilmozzi 2003). This massive cluster was selected because proper motions for several thousands stars were available in the literature. Combining proper motions with radial velocity information one gets all three components of the velocity vector, thus fully addressing the possible effects of anisotropy. Results, shown in Fig. 1, speak for themselves. The velocity dispersion profiles as derived for the three components of motion are very similar showing the cluster is isotropic (as is, by the way, also indicate by its very nearly circular shape). Moreover, the dispersion is found to be constant at large radii.

Following this initial result that, after all, could well be due to some effect like tidal heating, we were able to collect data for three more globular clusters: NGC 7078 (M15) and NGC 6171 (Scarpa, Marconi, and Gilmozzi 2004A,B), and NGC 7099 (Scarpa et al. 2006). Data for a fourth cluster, NGC 6341 (M92), appeared recently on the literature (Drukier et al. 2006) and also this cluster is presented here.

In all cases (Fig. 2) it was found that the velocity dispersion profile mimics precisely what is observed in high surface brightness elliptical galaxies. That is, the dispersion is maximal at the center, then decreases to converge toward a constant value at large radii where the acceleration goes below $a_0$.

With data for five clusters one can start comparing results and, interestingly, we found that in all cases the flattening of the dispersion profile occurs for very similar values of the internal acceleration of gravity. Values are collected in table 1, where we present for each cluster its total V magnitude, the radius where the flattening occurs, and the corresponding acceleration derived assuming a mass-to-light ratio of one. Within errors all profiles flattens out at $a_0$. It is worth noticing that these five clusters are all different in size, mass, position in the halo of the Milky Way, and dynamical history. Therefore, there is no obvious reason why they should conspire to have a similar velocity dispersion profile at large radii.

Galaxies provide us with another powerful tool to further disentangle non-Newtonian effects from other more classical phenomena like tidal heating (or a combination of effects like the distribution of dark remnants + tidal heating + cluster evaporation and so on) that might be responsible for increasing the velocity dispersion in the outskirt of globular clusters. High surface brightness galaxies (HSB) and low surface brightness galaxies (LSB) are known to behave differently. The latter having a remarkably flat velocity dispersion profile (e.g., Mateo 1997; Wilkinson 2006) allegedly due to LSB galaxies being dark-matter-dominated all the way to their center, while HSB are baryons-dominated at the center.

In view of what we have found for dense globular clusters that probe the same accelerations of HSB galaxies, it is natural to wonder whether low-concentration globular clusters behave like LSB galaxies. That is, do low-concentration globular clusters have constant velocity dispersion?

| Cluster Name | $M_v$ | R (pc) | a (cm s$^{-2}$) |
|---|---|---|---|
| NGC 5139 (ω Centauri) | -10.29 | 27±3 | 2.1±0.5×10$^{-8}$ |
| NGC 6171 (M107) | -7.13 | 8±2 | 1.3±0.6×10$^{-8}$ |
| NGC 6341 (M92) | -8.20 | 12±2 | 1.5±0.6×10$^{-8}$ |
| NGC 7078 (M15) | -9.17 | 20±2 | 1.4±0.4×10$^{-8}$ |
| NGC 7099 (M30) | -7.43 | 10±2 | 1.1±0.4×10$^{-8}$ |

The case of NGC 288.

In an attempt to answer this question we studied NGC 288, a low-concentration cluster located at 8.3 kpc from the sun and 11.6 kpc from the Galactic center.
With total absolute magnitude in the V band of M=-6.60 and central surface brightness of 19.95 mag arcsec$^{-2}$, NGC 288 has internal acceleration of gravity every where below $a_0$, as is the case for LSB galaxies. The initial selection of targets around NGC 288 was based on color, as derived from the analysis of ESO Imaging Survey frames. A catalog of targets was prepared including mostly stars from the sub-giant branch down to the turn off, between 15 and 18 apparent V mag. Observations were then obtained with FLAMES at the ESO VLT telescope. FLAMES is a fiber multi-objects spectrograph, allowing the simultaneous observation of up to 130 objects. We selected the HR9B setup that includes the magnesium triplet covering the wavelength range 5143 <λ< 5346 Angstrom at resolution R=25900. Stellar astrometry was derived cross correlating the stellar positions on the EIS frames with coordinates from the US Naval Observatory (USNO) catalog, which proved to have the required accuracy (0.3 arcsec) for FLAMES observations. Two different fiber configurations were necessary to allocate all the selected stars. For each configuration three 2700 s exposures were obtained under good atmospheric condition (clear sky and seeing ~1 arcsec) on August 29 and 30, 2005.

Radial velocities were derived cross correlating the spectra of each target with respect to a template, the target with the best spectrum. The two configurations shared a small number of stars, to evaluate and eliminate possible offsets in the velocity zero point. A posteriori, we verified that no correction was necessary down to a level of accuracy of 250 m s$^{-1}$, well below the accuracy required for our study. Finally, keeping in mind that we are interested only on the velocity dispersion, the global velocity zero point was derived by identifying few lines in the spectrum of the template.

As a whole, 126 radial velocities with accuracy better than 1 km s$^{-1}$ were derived. Of these, all but two were found to be cluster members, consistent with the very low contamination expected at the high galactic latitude (b=-89 degrees) of this cluster. To better constrain the velocity dispersion close to the cluster center, we combined our data with data for 24 additional stars mostly within 6 pc from the cluster center and radial velocity accuracy better that 1 km/s (Pryor et al. 1991). After applying an offset of 2.9 km/s to match our radial velocity zero point, these data smoothly merge with ours in the region of overlap, showing basically the same velocity dispersion (Fig. 3).

This combined sample was used to detect evidence for ordered rotation in NGC 288 that might contribute to sustain the cluster, finding no evidence for rotation down to the level of 0.5 km/s.

Velocities from this combined dataset allowed us to build a well sampled velocity dispersion profile from the center to almost 18 pc (Fig. 4). In Table 2 we report the velocity dispersion in km/s with 1$\sigma$ uncertainties, together with the limits of the bins, the number of stars in each bin, and the bin center define as the average of the radii of the stars in the bin.

Looking at both Fig. 3 and 4, we see no indications of a vanishing velocity dispersion at large radii, rather, the dispersion is remarkably constant, being consistent with the average value of 2.3±0.15 km/s over the full range of radii covered by the data.

The five clusters ω Cen, NGC 7078, NGC 6171, NGC 7099, and NGC 6341 while having different sizes, different masses, and different dynamical histories, have the common property of being highly concentrated, thus that the acceleration of gravity in their central regions is above $a_0$. Only in the outskirts the acceleration is below this value. All these clusters are found to behave like HSB elliptical galaxies having constant velocity dispersion at large radii. By contrast, NGC 288 has the ``peculiar'' property of being rather diffuse, with a central surface brightness of ~20 mag arcsec$^{-2}$ in the V band. This has the important consequence that the internal acceleration of gravity is extremely small. Indeed it can be shown to be everywhere below $a_0$ for any of the typical mass-to-light ratio assumed for globular clusters.

It is well known that LSB do have velocity dispersion profiles remarkably flat with no central maximum (Mateo 1997; Wilkinson et al. 2006).

If this is due to a breakdown of Newtonian dynamics below $a_0$ (and not because of dark matter), then the velocity dispersion of NGC288 should mimic what is observed in these galaxies. Within errors, this is the certainly the case (Fig. 4).

The similarity between NGC 288 and LSB galaxies is striking. The canonical explanation for the constant velocity dispersion in LSB galaxies is that these objects are dark-matter-dominated all the way to their center.

**Table 2.** Radial velocity dispersion of NGC 288

| Bin limits (pc) | Stars/bin | bin center (pc) | $\sigma$ (km s$^{-1}$) |
|---|---|---|---|
| 0 – 2 | 8 | 1.34 | 2.32 ± 0.64 |
| 2 – 4 | 21 | 3.21 | 2.41 ± 0.41 |
| 4 – 6 | 20 | 5.09 | 2.73 ± 0.46 |
| 6 – 8 | 22 | 7.18 | 2.52 ± 0.41 |
| 8 – 10 | 20 | 9.02 | 2.60 ± 0.45 |
| 10 – 12 | 14 | 10.71 | 2.16 ± 0.46 |
| 12 – 14 | 25 | 12.95 | 1.90 ± 0.31 |
| 14 – 20 | 17 | 16.12 | 2.17 ± 0.42 |

In the case of a globular cluster this explanation is quite unpalatable. Thus, looking at both our results for globular clusters and the amazing ability of a particular modification of the Newtonian Dynamics known as MOND (Milgrom 1983) to describe successfully the properties of a large number of stellar systems without invoking the existence of non-baryonic dark matter, we see here evidence for a failure of Newtonian Dynamics below $a_0$.

Given the relevance of this claim, we urge the astronomical community to disprove or generalize our results.

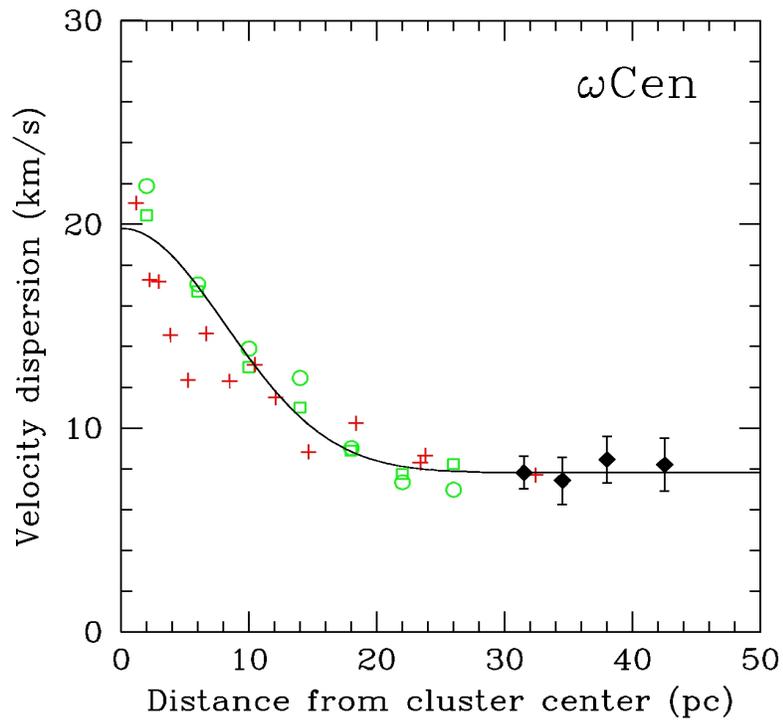

Fig. 1. The velocity dispersion profile of ω Centauri. Circles and squares represent the dispersion as derived from proper motion data. Crosses are radial velocity dispersions from the literature, to which we added data for 75 stars (the four last points with error bar). The solid line is not a fit to the data. It is a Gaussian plus a constant drawn to emphasize the flattening of the dispersion at large radii.

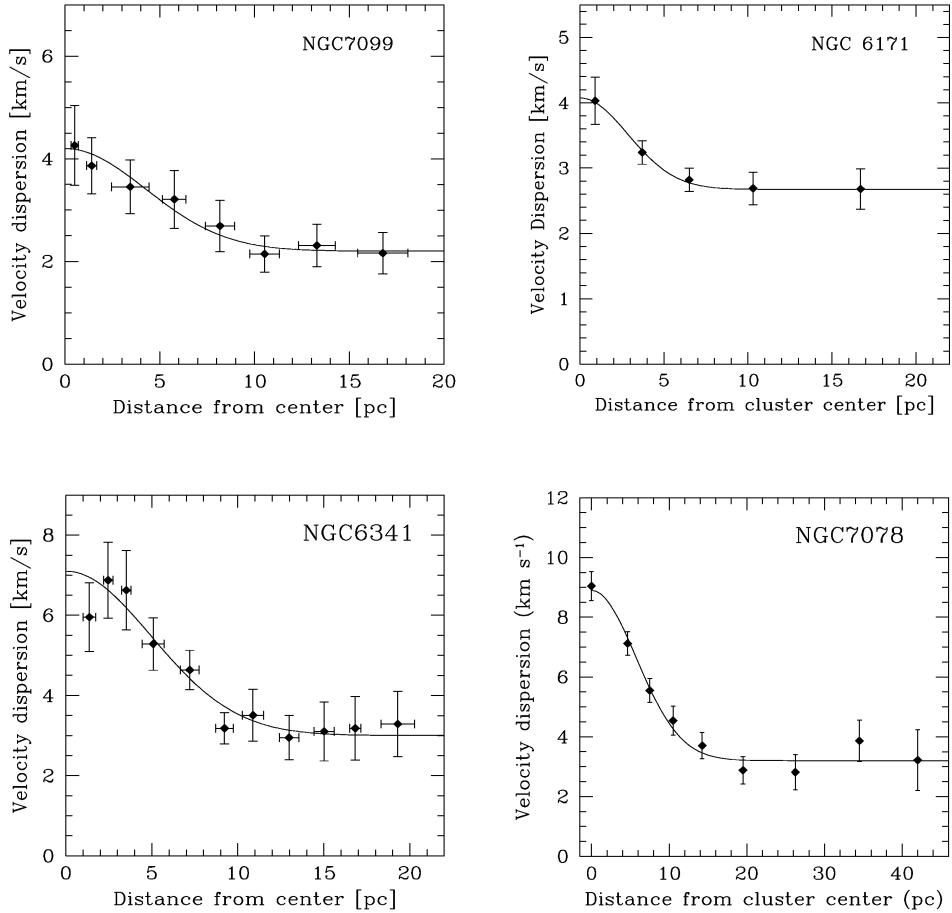

Fig. 2. Velocity dispersion profiles of the remaining four clusters studied so far. In all cases the velocity dispersion is maximal at the center, then decreases to converge toward a constant value at large radii. This is typical of high surface brightness elliptical galaxies. Note the profile of NGC 6171 and NGC 7099 were derived from our own data, while the profile of NGC 7078 and NGC 6341 were both derived from data found in the literature (Drukier et al. 2006 and reference therein). The solid line is not a fit to the data. It is a Gaussian plus a constant drawn to emphasize the flattening of the dispersion at large radii.

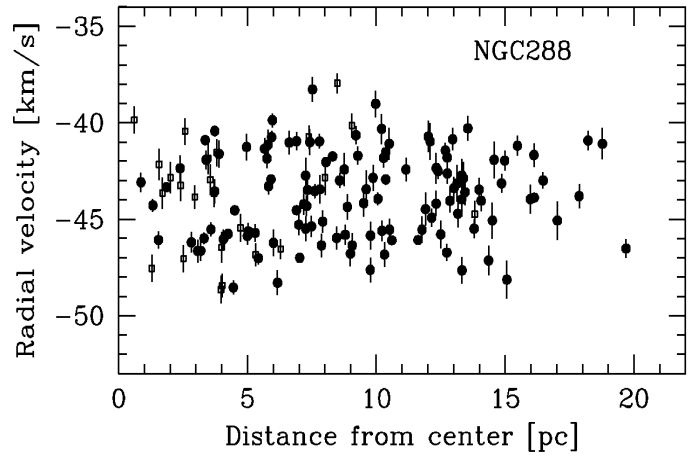

Fig. 3. Radial velocity distribution for the low-concentration cluster NGC288. The 124 stars studied as part of this work are shown as solid squares, while 24 velocities from Prior et al. 1991 are shown as open squares.
Data distributes uniformly from the center to 20 pc, showing no indication of a decrease of the dispersion.

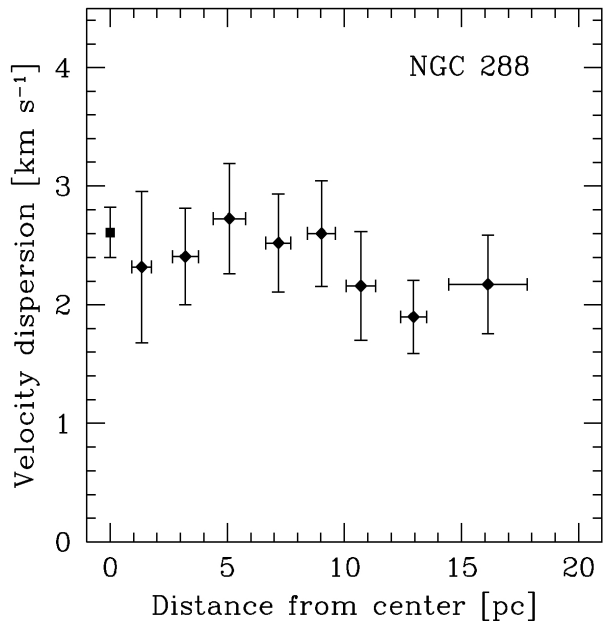

Fig. 4. The velocity dispersion profile of the low concentration cluster NGC 288. From the point of view of the internal accelerations of gravity this cluster is the equivalent of a low surface brightness galaxy and, similarly to what is found in these galaxies, the dispersion profile is remarkably flat.